\documentclass[iop,twocolappendix]{emulateapj}
\usepackage{natbib}
\usepackage{float}
\usepackage{times}
\usepackage{url}
\usepackage[left=2.5cm,right=1.0cm,top=4.0cm,bottom=0.0cm]{geometry}

\usepackage{xcolor}

\usepackage{amsmath,amssymb,amsfonts}
\usepackage{graphicx}
\usepackage[breaklinks,colorlinks,citecolor=blue,urlcolor=blue]{hyperref}
\usepackage[all]{hypcap}

\slugcomment{Accepted for publication in The Astrophysical Journal}

\shorttitle{Measurement of mass segregation}
\shortauthors{Yu {\it et~al.}}

\begin{document}

\title{Simulations of Fractal Star Cluster Formation: I.~New Insights for Measuring Mass Segregation of Star Clusters with Substructure}

\author{Jincheng~Yu$^{1,2}$, Thomas H.~Puzia$^{1}$, Congping~Lin$^{3,4,5}$, and Yiwei~Zhang$^{3,4,5}$}
\affil{
$^{1}$Institute of Astrophysics, Pontificia Universidad Cat\'olica de Chile, Av.~Vicu\~na Mackenna 4860, 7820436 Macul, Santiago, Chile; \href{mailto:yujc.astro@gmail.com}{yujc.astro@gmail.com}, \href{mailto:tpuzia@gmail.com}{tpuzia@gmail.com}\\
$^{2}$National Astronomical Observatories, Chinese Academy of Sciences, 20A Datun Road, Chaoyang District, Beijing 100012, China\\
$^{3}$Center for Mathematical Science, Huazhong University of Science and Technology, 1037 Luoyu Road, Wuhan 4370074, China; \href{mailto:congpinglin@gmail.com}{congpinglin@gmail.com}, \href{mailto:yiweizhang831129@gmail.com}{yiweizhang831129@gmail.com}\\
$^{4}$Hubei Key Laboratory of Engineering Modeling and Scientific Computing, Huazhong University of Science and Technology, 1037 Luoyu road, Wuhan 430074, China\\
$^{5}$School of Mathematics and Statistics, Huazhong University of Science and Technology, 1037 Luoyu Road, Wuhan 4370074, China
}

\begin{abstract}
	We compare the existent methods including the minimum spanning tree based method and the local stellar density based method, in measuring mass segregation of star clusters. We find that the minimum spanning tree method reflects more the compactness, which represents the global spatial distribution of massive stars, while the local stellar density method reflects more the crowdedness, which provides the local gravitational potential information. It is suggested to measure the local and the global mass segregation simultaneously. We also develop a hybrid method that takes both aspects into account. This hybrid method balances the local and the global mass segregation in the sense that the predominant one is either caused either by dynamical evolution or purely accidental, especially when such information is unknown a priori. In addition, we test our prescriptions with numerical models and show the impact of binaries in estimating the mass segregation value. As an application, we use these methods on the Orion Nebula Cluster (ONC) observations and the Taurus cluster. We find that the ONC is significantly mass segregated down to the 20th most massive stars. In contrast, the massive stars of the Taurus cluster are sparsely distributed in many different subclusters, showing a low degree of compactness. The massive stars of Taurus are also found to be distributed in the high-density region of the subclusters, showing significant mass segregation at subcluster scales. Meanwhile, we also apply these methods to discuss the possible mechanisms of the dynamical evolution of the simulated substructured star clusters.
\end{abstract}

\keywords{globular clusters: general -- methods: numerical -- methods: data analysis}

\section{Introduction}
\label{sec:introduction}

Most stars are considered to be forming in star clusters or associations, which disperse thereafter on timescales proportional to the mass of the initial cluster/association \citep[e.g.][]{kru12}. However, the early formation and evolution of star clusters is a complex problem spanning a vast physical scale range and is thus still not well understood \citep[e.g.][]{bat03, bat14, ren15}. An interesting and widely discussed aspect of the star cluster formation process is the progression of mass segregation, which generally refers to the state in which massive stars are more centrally concentrated, i.e.~the massive-star subpopulation in a cluster is more compact as a group and is also located in a denser region.

Many different measurement tools exist to determine the degree of mass segregation in star clusters, including comparing the slope of the mass function (either differential or cumulative) and the characteristic radii, e.g.~the half-mass radii of various stellar subpopulations with different mass ranges. However, these methods all assume a spherical star configuration, while many young star clusters are observed to be substructured \citep[e.g.][]{hil97, fue06, chen07, wang08}. To overcome the complication of nonsphericity, \cite{all09a} proposed measuring mass segregation based on the minimum spanning tree (MST) method. Given a set of vertices, a spanning tree connects all the reference points (stellar positions) together without forming loops. The lines in between the vertices are generally referred to as edges. Consequently, the MST is then a spanning tree with minimum length. Generally, an MST is constructed for a chosen group of stars and then its length is calculated as a proxy of the segregation indicator when compared to random star sets \citep[see][for details]{all09a}.

\begin{figure*}
	\centering
	\includegraphics[width=0.85\linewidth,angle=0]{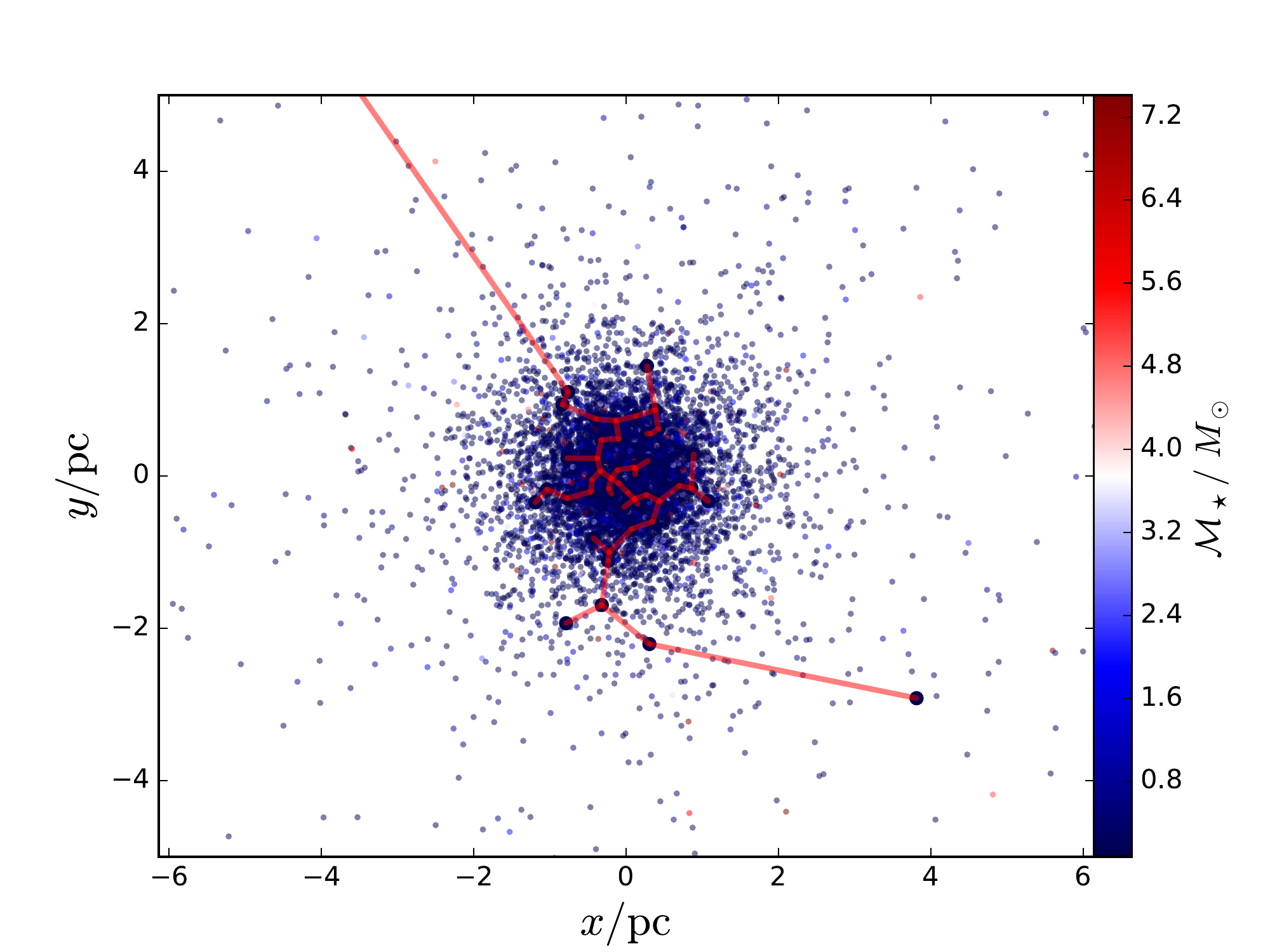}
	\caption{Illustration of the stellar distribution in a numerical simulation of a spherically symmetric globular cluster. The plot shows individual stellar particles that have color shading parametrized by their mass and indicated by the color bar. The 50 most massive stars in the simulation are shown as large dots (correspondingly shaded), and their positions are connected by their MST.~Note the long edge toward an outlier at the top left.}
	\label{fig:mst_illustration}
\end{figure*}

The MST method was shown to be efficient and effective in dealing with clumpy distributions. However, \cite{par11} found an ``inverse'' mass segregation for the Taurus star-forming region using the MST method \citep{mas11}. As the MST length is the summation of all the MST edges, one outlier may greatly increase the MST length (see Figure~\ref{fig:mst_illustration}). Irrespective of such outliers, their presence should indeed have a much smaller effect on the compactness of the group. Note that the MST edge length distribution may be asymmetric, which is why using the MST length, or, equivalently, the average of MST edge length, appears no longer suitable to represent the compactness of a certain set of stars. \cite{olc11} also notice this flaw, and they introduced a variant based on the geometric average of the MST edges to improve Allison's original MST method.~\cite{mas11} suggest to use the median MST edge length in a similar spirit. In general, using a geometric average or median MST edge can improve the performance.

However, we note that the methods based on the MST consider the information of compactness without giving any information on the role that massive stars play in their distribution as a group. That is, when a star cluster is measured to be mass segregated via the MST methods, the massive stars should be generally close to each other. But these stars can also be grouped in a clump that is relatively far away from the other stars in the cluster, i.e.~a subclump of massive stars that is relatively far away from the cluster center of mass (see Figure~\ref{fig:mst_lnnd1}). In this case, if other clumps are significantly denser than the clump, which most massive stars are, the massive stars should not be considered as centrally concentrated, although this kind of distribution can still be claimed as ``mass segregation.'' However, a more robust definition of mass segregation that can indicate the dynamical status of the star cluster would be more helpful rather than a definition purely based on geometrical arguments. Therefore, massive stars should also be distributed in a high-density region if the star cluster is to be defined as mass segregated. The local stellar density of a massive star is a good starting point to determine its surrounding density distribution. \citeauthor{mas11} use the local stellar number density to measure the degree of mass segregation. On the other hand, if the massive stars are distributed in high-density regions, they may disperse in many different clumps that are relatively far away from each other. In this case, the MST methods describe the global status of massive stars. \cite{par15} show that  measuring mass segregation using MST methods and the local stellar density method yield different definitions, and they suggest to use these two methods in tandem.

In this paper, we use the Local $n$th Neighbor Distance (LnND) as an indicator of the crowdedness for a group of stars and use the MST edge as a proxy of compactness. We show that the LnND gives the local information while the MST edge gives the global information. Both indicators should be measured simultaneously in order to better describe the mass segregation of star clusters. We also introduce a new hybrid mass segregation parameter that contains both global geometric information and local potential information by combining the compactness indicator and the crowdedness indicator. The paper is structured as follows. In Section~\ref{sec:methods} we describe the MST and the LnND methods, as well as our hybrid method of measuring the degree of mass segregation, and discuss some additional caveats along the way. We test these methods with an $N$-body simulation and observational data and discuss their differences in Section~\ref{sec:discussion}, and we conclude this work in Section~\ref{sec:conclusions}.

\section{Methods}
\label{sec:methods}

\subsection{Measurements of mass segregation}
\label{subsec:mass_segregation}
We aim at measuring the mass segregation of very young star clusters. The origin of mass segregation of young star clusters is of crucial importance to determine the star formation scenario \citep[e.g.][]{bon01, wei06}. However, many traditional methods, including comparing the slope of the mass function and the characteristic radii, deal with a spherical star cluster configuration, while the star clusters are born with substructures. To overcome the complication of nonsphericity, the MST method \citep[e.g.][]{all09a, olc11, mas11} and the local stellar density method \citep[e.g.][]{mas11} are implemented to measure the mass segregation of young star clusters.

Both methods can be used to measure the mass segregation of such substructured star clusters. However, they show subtle differences since they measure the different aspects of mass segregation. The MST method in essence measures the compactness of a group of stars, while the LnND method measures the crowdedness. The compactness of a group of stars refers to how close these stars are. If the stars in a group are located very close to each other, then they are highly compact, which we define as high compactness. On the other hand, the crowdedness of a star refers to their proximity. If the chosen star is located in a very dense region, then it is highly crowded, which we define as high crowdedness. In general, the compactness gives the global geometric information, while the crowdedness gives the local potential information.

\subsubsection{Measuring compactness with the minimum-spanning-tree (MST) methods}
We first compare the different MST methods: (1) the geometric average of the MST edge (gmMST) from \cite{olc11} and (2) the median MST edge (mMST) from \cite{mas11}. These MST methods measure the degree of mass segregation by comparing the compactness of massive stars and the compactness of the whole cluster. The compactness of the whole star cluster is derived from a series of groups of randomly chosen stars that all contain the same number of stars as the massive-star group. With a large enough number of randomly chosen groups, their average or median value is a good estimate of the whole star cluster, while the dispersion can be used as a measure of the uncertainty. The degree of mass segregation measured by the MST methods is typically written as
\begin{equation}
	^m{\Lambda} = \frac{^{m}\tilde{l}_{\rm random}} {^{m}l_{\rm massive}},
	\label{eqn:mst}
\end{equation}
where $m$ is the number of massive stars, while $l$ is the geometric mean or median of the MST edges of the chosen star group, and $\tilde{l}_{\rm random}$ is the median of $l_{\rm random}$. $^m{\Lambda} > 1$ means that the star cluster is mass segregated down to the $m$th most massive star. If, on the other hand, $^m{\Lambda}=1$ then the massive-star sample is considered to be not mass segregated. These different MST approaches are more robust against outliers than Allison's original method, which uses the total length of the MST and is equivalent to the arithmetic average of the MST edge.~In Figure~\ref{fig:mst_illustration} we show a star cluster containing 10,000 stars with its stellar mass randomly distributed following a Plummer density profile so that it is not mass segregated. We populate the stellar content according to the \cite{kro02} initial mass function (IMF) within the stellar mass range $0.08\!-\!100\, M_\odot$. The mass segregation values derived from the geometric average and median MST edge method show that there is no mass segregation in the model setup (i.e. gmMST=0.94 [0.78, 1.17] and mMST=1.01 [0.91, 1.12], where the numbers in brackets indicate the 25th and 75th percentiles, respectively). We use the Delaunay triangulation (2D) and the $k$-d tree technique (3D) to accelerate the calculation of the MST edges, as discussed in the Appendix.

\subsubsection{Measuring crowdedness with the LnND method}
We use the distance of the sixth-nearest star around a chosen star to estimate its local density \citep{vHoe63, cas85}. The local density of a chosen star is one-to-one correspondent with $r_{6}$, the sixth-nearest star distance. For example, in a 2D case, the local density satisfies
\begin{equation}
	{\Sigma} = \frac{6 - 1} {\pi r_{6}^{2}},
\end{equation}
Previous works have used this quantity to measure the mass segregation by defining a ratio between the local surface density of a subset of massive stars and randomly chosen star sets \citep[e.g.][]{par15}.

Instead of using the local surface density (2D case), we use the LnND for each star, which is a measure that has the same dimensions as the MST edge. Note that the smaller the LnND becomes, this naturally means that the star is in a higher local density environment. Hence, if a group of massive stars has a smaller LnND, it is more likely to be mass segregated. The degree of mass segregation measured by the LnND method is similar to that measured by the MST edge methods given in Equation~\ref{eqn:mst}:
\begin{equation}
	^m{\Sigma} = \frac{^{m}\tilde{d}_{\rm random}} {^{m}d_{\rm massive}},
	\label{eqn:lnnd}
\end{equation}
where $m$ is the number of massive stars, while $d$ is the median of $r_6$ of a chosen star group, and $\tilde{d}_{\rm random}$ is the median of $d_{\rm random}$. $^m{\Sigma}\!>\!1$ means that the star cluster is mass segregated down to the $m$th most massive star. Similar to $^m{\Lambda}$ discussed before, $^m{\Sigma}\!=\!1$ means no mass segregation for the corresponding massive-star sample. We note, however, that other (e.g., less massive) stars can have other MST and LnND measures and, therefore, not necessarily be similarly mass segregated.

The LnND gives the crowdedness of a star, hence providing us with the local potential information, which is essential in discussing the genesis of mass segregation. We utilize the KNN algorithm to determine the sixth-nearest star, which requires $\mathcal{O}\bigl(k \log N \bigr)$ of computation time, as discussed in the Appendix. We calculate the degree of mass segregation in the star cluster of Figure~\ref{fig:mst_illustration} using the LnND method and get a close to unity value (i.e.~$^m{\Sigma}=1.03\,\,[0.94, 1.14]$, where the numbers in the brackets are the 25th and 75th percentiles, respectively), which is consistent with the model setup.

\subsubsection{Comparison of MST and LnND: Treatment of special vertex configurations}
\label{sec:mst_vs_lnnd}

\begin{figure*}
	\centering
	\includegraphics[width=0.75\linewidth,angle=0]{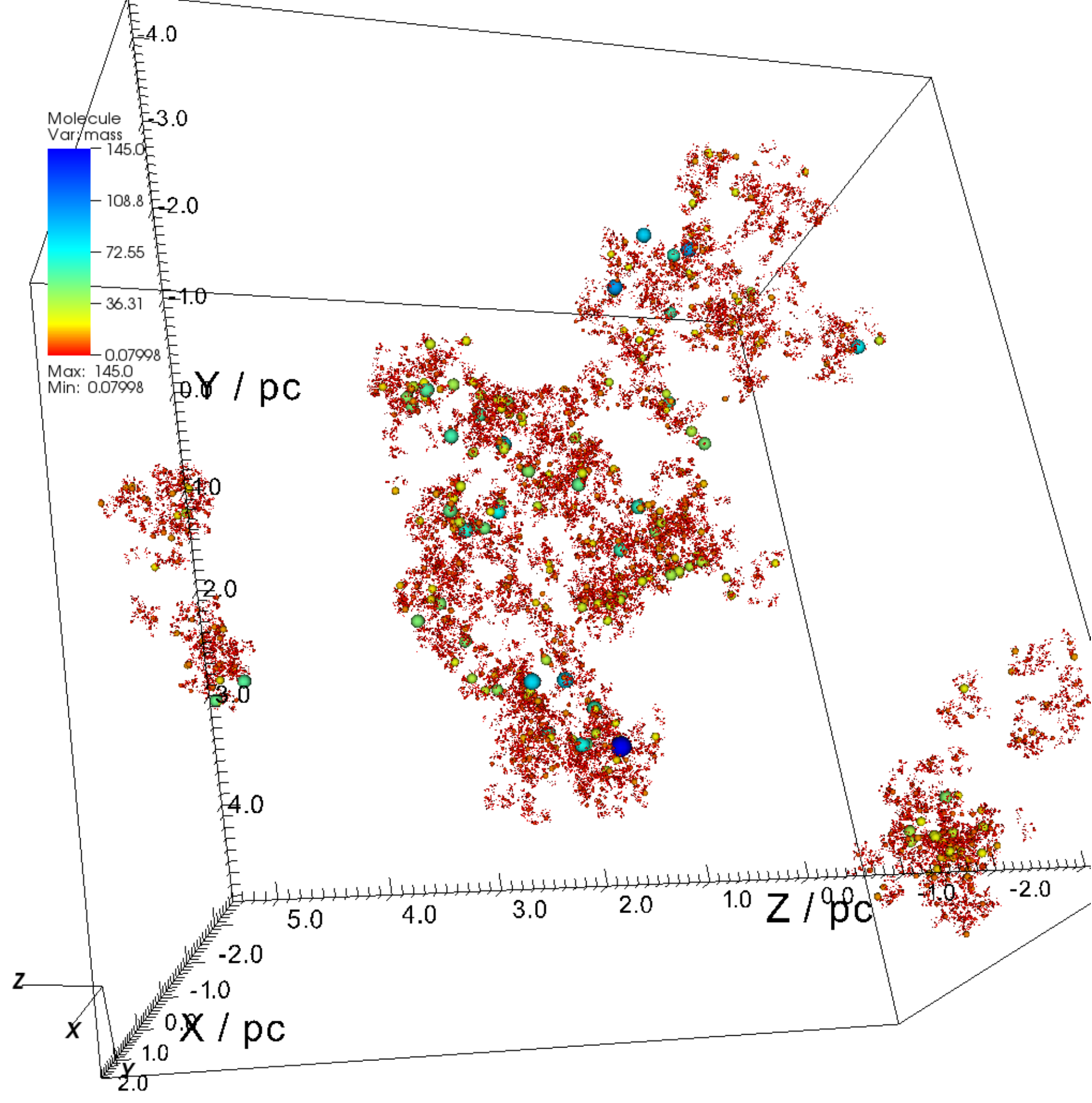}
	\caption{Illustration of the initial star configuration of one of our simulated fractal star clusters. The fractal stellar distributions are used to represent the substructured distribution of observed star forming complexes, e.g.~the Carina star-forming region \citep[e.g.][]{zei16} and the Tarantula Nebula \citep[e.g.][]{sab16}. The star cluster has several ``main'' clumps, each of which has its own substructure. Stars are colored according to their mass and are randomly distributed throughout the structure, i.e.~there is no preference where massive stars are initially placed.}
	\label{fig:init}
\end{figure*}

The MST method and the LnND method are shown to be effective in measuring mass segregation of substructured star clusters. The MST method measures the compactness of a group of stars, while the LnND method measures the crowdedness. In general, these two methods are consistent with each other. In a mass-segregated star cluster, massive stars are very likely to be close to each other and located in a denser region. However, a compact group of massive stars does not necessarily mean that their local density is high. The discrepancy becomes more essential when the star cluster is substructured.

Consider a fractal star cluster of 100,000 stars, shown in Figure~\ref{fig:init}. The massive stars are shown as blue spheres, while the low-mass stars are shown as red spheres. The masses of the stars are randomly assigned so that the star cluster is set to be non-mass-segregated. Note that if the masses of the stars are carefully assigned so that the star cluster becomes mass segregated, there will be two possibilities. One possibility is that the massive stars (blue spheres) are set to be dispersed in the central part of each clump. In this case, the LnND method would give positive results because the massive stars are located in relatively high density regions, while the MST method would fail to detect the mass segregation because the distances between massive stars do not show significant differences from the other stars. Alternatively, most of the massive stars (blue spheres) are set to be located in very few clumps or even one clump. In this case, the MST method would detect mass segregation because the massive stars are very close to each other, while the LnND method would show non-mass-segregated result because the local density of each clump is similar, due to the self-similarity provided by the properties of the fractal distribution.

\begin{figure*}
	\centering
	\includegraphics[width=0.8\linewidth,angle=0]{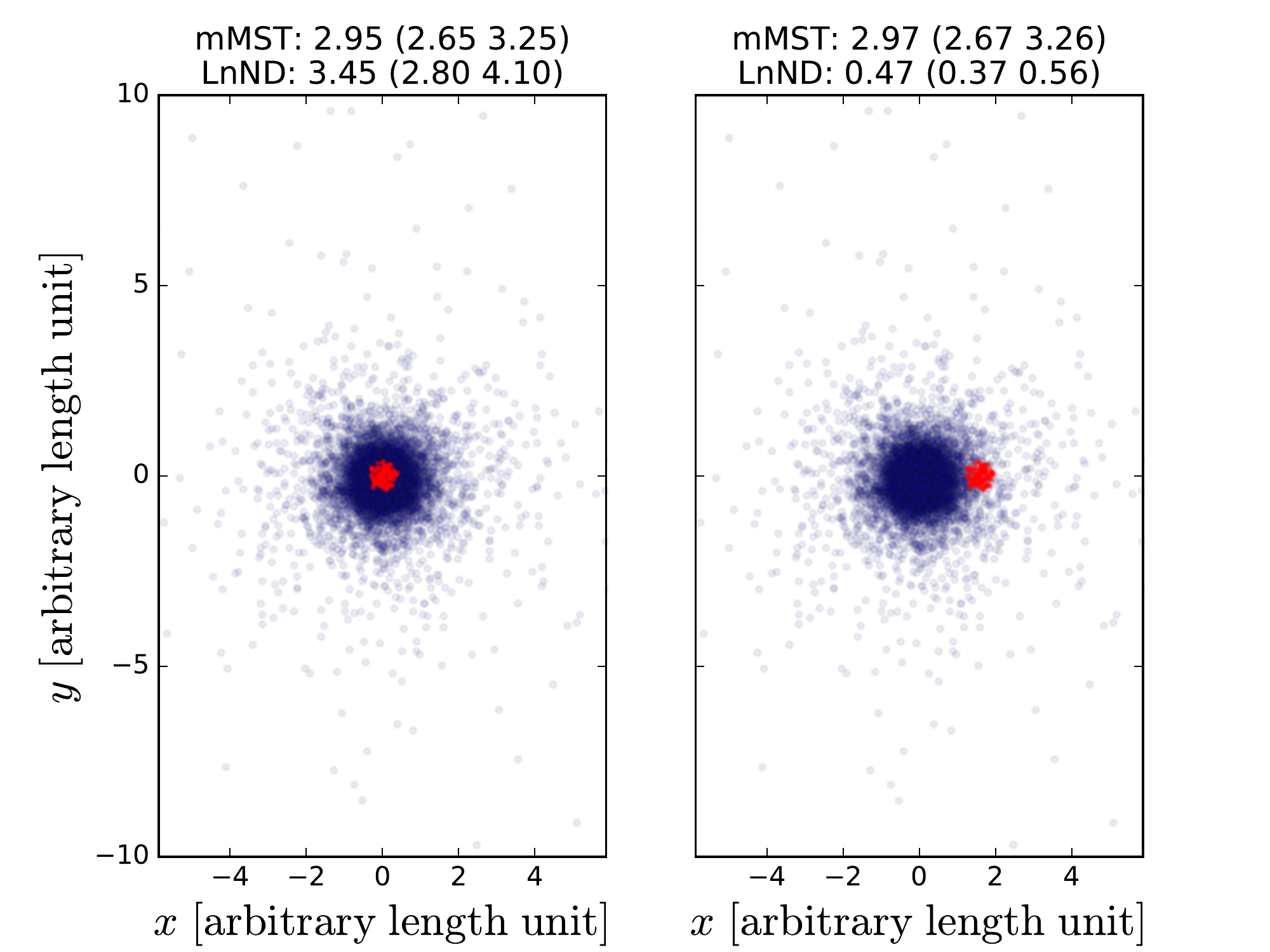}
	\caption{Left panel: Plummer model of a star cluster, with the top 50 massive stars (red circles) located in the inner (dense) region. Right panel: Same model as in the left panel, but with massive stars shifted to the outer regions of the cluster. The degree of mass segregation for the 50 most massive stars, with 25th to 75th percentiles, measured by different methods is indicated at the top of each panel. Note that the MST method gives almost the same degree of mass segregation, while the LnND method suggests that these two configurations differ.}
	\label{fig:mst_lnnd1}
\end{figure*}

We point out that the MST method itself only gives the compactness of massive stars, while it does not give any information about whether the massive stars are in the dense region of the cluster or not.  To further illustrate this, instead of using the fractal distributed star cluster (Figure~\ref{fig:init}), we use an artificial configuration with a Plummer density distribution of 10,000 stars, shown in Figure~\ref{fig:mst_lnnd1}. We do not include any binaries in order to make our test as simple as possible. The 50 most massive stars are moved to the cluster central regions, defined as a spherical region within 50\% of the half-mass radius, $r_{\rm hm}/2$, so that the star cluster is highly mass segregated (left panel of Figure~\ref{fig:mst_lnnd1}). The MST method gives for this configuration positive results. We then shift these massive stars to the outer regions of the cluster without changing the location of the other stars, as shown in the right panel of Figure~\ref{fig:mst_lnnd1}. Massive stars are then not as highly mass segregated as the original distribution. However, the MST method still gives almost the same mass segregation degree, because the compactness of massive stars remains, while the LnND method clearly reflects the changes. If a star cluster is clumpy, massive stars can be relatively far away from the majority of the other stars. Such a configuration, although unrealistic, clearly poses problems to the accuracy of the MST method since the MST method focuses on the compactness of massive stars. In a more realistic configuration, e.g.,~substructured configuration (see Figure~\ref{fig:init}), if the massive stars happen to appear in one clump, a similar problem occurs if the mass segregation is measured by the MST method.

\begin{figure*}
	\centering
	\includegraphics[width=0.8\linewidth,angle=0]{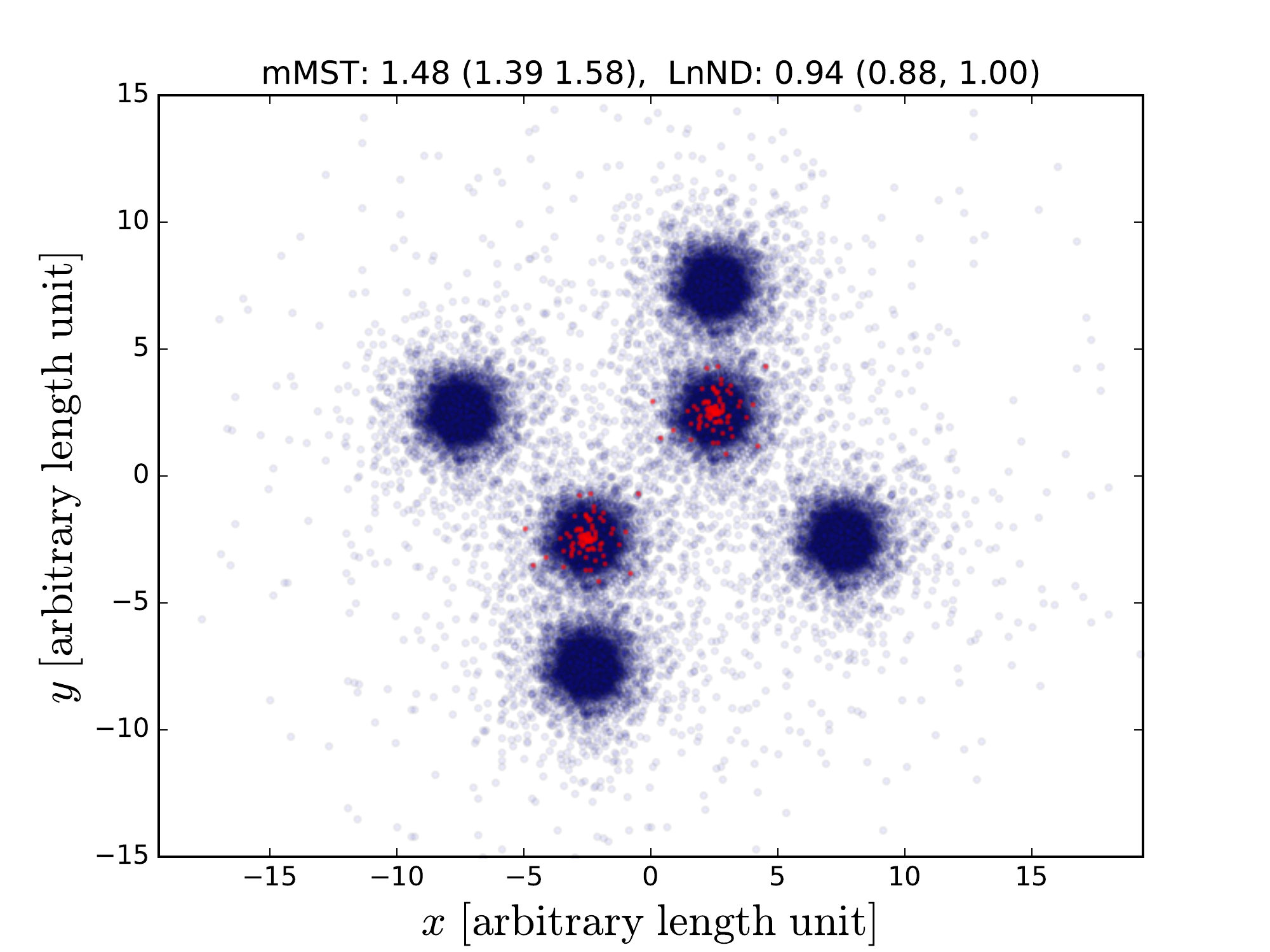}
	\caption{Clumpy model of a star cluster with six subclusters. Only the inner two clumps contain massive stars (red circles). See text for details. The degree of mass segregation for the 100 most massive stars, with 25th to 75th percentiles, measured by the different methods, i.e.~mMST (Eqn.~\ref{eqn:mst}) and LnND (Eqn.~\ref{eqn:lnnd}), is indicated at the top of the panel.}
	\label{fig:mst_lnnd2}
\end{figure*}

On the other hand, the LnND method itself may also be ambiguous in some special cases. To better illustrate this, we construct an idealized star cluster with six subclusters, while each clump follows a Plummer distribution with 10,000 stars, shown in Figure~\ref{fig:mst_lnnd2}. Such a configuration allows us to analyze the mass segregation both locally, i.e.,~the mass segregation of each clump, and globally, i.e.,~the whole star cluster, since the bias from the cluster identification is negligible. Massive stars are carefully set to be only located in the two inner clumps. Both of the inner clumps contain 50 massive stars, but within the clump massive stars have the same distribution as the low-mass stars, so that the cluster is globally mass segregated, while being non-mass-segregated locally. In this scenario, the LnND method gives a non-mass-segregated result, which is exactly the same as the mass segregation condition within each clump. Therefore, the LnND only gives the local mass segregation status. In contrast, the MST method clearly gives the global result that the star cluster is mass segregated. This configuration shows that the LnND method focuses on the crowdedness of massive stars. In a more realistic configuration, e.g.,~substructured configuration (see Figure~\ref{fig:init}), if the massive stars happen to appear in ``inner'' clumps, a similar problem occurs if the mass segregation is measured by the LnND method.

As \cite{par15} pointed out, the MST method measures mass segregation where the massive stars are concentrated, while the LnND method measures the mass segregation where massive stars are located in a high-density region. Furthermore, in a multiclump star cluster, the MST method reflects more the global mass segregation state of the system, while the LnND method reflects more the local mass segregation.

\subsection{Hybrid mass segregation indicator}
Based on the previous insights, neither the compactness, measured by the MST method, nor the crowdedness, measured by the LnND method, accurately describes the mass segregation of star clusters alone. It is important to measure the mass segregation of star clusters with both indicators. In addition, it is also useful to describe the degree of mass segregation with one parameter. We therefore combine these two methods and give a hybrid method. The MST median edge of the $m$ most massive stars is used as their compactness indicator, while the median value of the LnND of the same set of massive stars, defined as the distance to the sixth-nearest star, is used as their crowdedness indicator. We calculate the geometric mean of these two values to define the new hybrid segregation factor
\begin{align}
	^{m}\zeta = {}& \left( {\rm med[MST] \times med[LnND] } \right)^{1/2},
\end{align}
where $m$ is the number of massive stars, and med[MST] and med[LnND] take the median value of the MST edges and the median of the LnND values for the star sample, respectively. Here we use the geometric mean of the MST and LnND measurements, given that they are of the same dimensions.

We then choose the same number of stars randomly from the cluster stellar population to calculate the $\zeta_{\rm random}$ for multiple times, and we take the median value $\tilde{\zeta}_{\rm random}$ as a reference, in the spirit of Allison's method. Finally, we give our new definition of the mass segregation degree,
\begin{equation}
	^m\hat{\zeta} =\, ^m\tilde{\zeta}_{\rm random}/^m\zeta_{\rm massive}.
	\label{eqn:zeta}
\end{equation}
Similarly, $^m\hat{\zeta} > 1$ means that the star cluster is mass segregated down to the $m$th massive star.

In general, if star clusters are mass segregated, they are mass segregated both globally and locally. However, star clusters, especially fractal or substructured star clusters, can be mass segregated in two different ways, either predominantly globally or locally. The key question is then how we define mass segregation of star clusters. The problem of mass segregation is normally considered as a result of star formation and dynamical evolution. If a substructured star cluster is born mass segregated, massive stars can be close to each other if massive stars are tend to form as a group; alternatively, massive stars can be only distributed in the denser region if they are formed in a higher-density region.

Similarly, from the dynamical evolutionary point of view, if the mass segregation of the young star cluster is caused by rapid dynamical evolution, these two assumptions still hold. One possible evolution picture is that star clusters are first globally mass segregated, while showing little local mass segregation, and then evolve to a state that has similar local and global mass segregation with the disappearance of substructures. Alternatively, star clusters can be first locally mass segregated, while showing little global mass segregation, and then evolve to a state that has similar local and global mass segregation. 

Moreover, knowing the mass segregation that is predominantly global or predominantly local does not mean knowing the evolutionary state of the star cluster. There has been no direct way to determine the differences of dynamical evolutionary states between a predominantly globally mass segregated star cluster and a predominantly locally mass segregated star cluster only by investigating the mass segregation. We keep both possibly valid scenarios by treating star clusters with similar $^m\hat{\zeta}$, the estimate of mass segregation with the hybrid method, as having similar degrees of mass segregation. Furthermore, we can trace the mass segregation evolution of such star clusters with these parameters. This is discussed in Section~\ref{sec:simulation} and will be discussed in detail in a subsequent paper of this series.

We conclude that while the MST gives us the compactness of massive stars and, therefore, their global mass segregation estimate, the LnND gives us their crowdedness and, therefore, their local mass segregation indicator. Combining these two schemes balances these two spatial aspects, both of which are necessary to determine the status of mass segregation in a star cluster with substructure. This also makes the quantification of the mass segregation for stellar systems without the presence of much substructure more robust.

\subsection{The impact of binary stars}
\label{sec:binary_impact}

\subsubsection{Binary parameters}
The situation becomes slightly more complex when binary stars are considered. To study their impact, we make the same model with the same total number of stars as shown in Figure~\ref{fig:mst_illustration}, but this time containing 50\% and 10\% of binaries. 

The initial orbit distribution of the binaries is determined by the period and eccentricity distribution function. We adopt a period distribution function of the following form:
\begin{equation}
	f_{P} = 2.5 \frac{\log_{10} P - \log_{10} P_{\rm min}} {45 + (\log_{10} P - \log_{10} P_{\rm min})^2},
	\label{eqn:bin_period}
\end{equation}
where $P$ is the period of the binary in days and $P_{\rm min}$ is the minimum period. With the standard convention, we set $\log_{10} P_{\rm min} = 1$.

A thermal distribution function is used as the distribution of eccentricity,
\begin{equation}
	f_{e}(e) = 2e,
\end{equation}
where $e$ is the eccentricity of the binary orbit \citep[see][for more details]{kro95a, kro95b}.

\subsubsection{Binary pairing}

\begin{figure*}
	\centering
	\includegraphics[width=0.8\linewidth,angle=0]{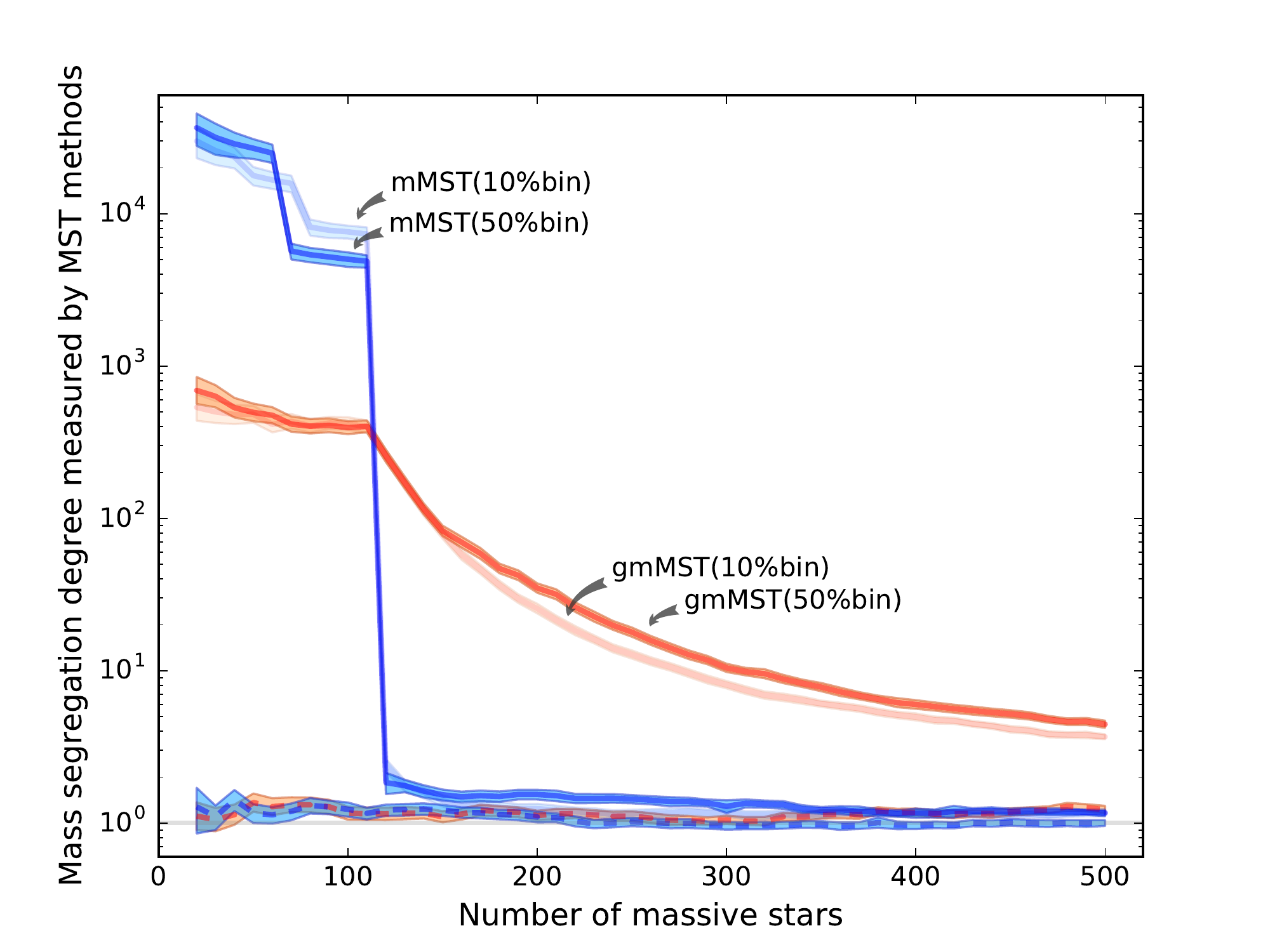}
	\caption{Mass segregation degree based on the geometric-mean MST method (gmMST; red curves) and the median MST method (mMST; blue curves) as a function of the massive-star group size, i.e. the number of most massive stars, for a star cluster with 50\% (dark shading) and 10\% binaries (light shading). Binary stars are treated here as resolved entities, where all stellar components contribute to the gmMST and mMST measurement. The curves for random binary pairing functions are shown by dashed lines, while ordered pairing functions are given by solid lines. Note that the curves for random binary pairing functions are close to unity (horizontal gray line). The shaded bands give the corresponding measured 25th to 75th percentiles.}
	\label{fig:methods_compare_nbr}
\end{figure*}

Another binary parameter is the stellar mass ratio distribution. The choice of the mass ratio can be called the pairing function, which is combining stars into binary systems. When random pairing is used (i.e.~binary companions are randomly chosen from a given IMF), we find similar mass segregation results to those for star clusters containing only single stars. However, random pairing is ruled out observationally, and there is also a lack of theoretical backing for a random pairing function \citep{sha02, kou05, kou07a, kou07b}. Indeed, massive stars preferentially choose other massive stars as their binary companions \citep{kou10}.~This implies that both the geometric average and the median MST method may be biased by the short MST edges from binary systems, although the LnND method is less influenced.~An alternative binary pairing function is the so-called ordered pairing \citep[i.e.~mass ratio of the binary star components $\sim\!1$, in particular for massive stars; see e.g.][]{kou09, oh12}, which matches stars in order of their mass distribution. We re-pair our test model using ordered pairing and find some extremely high values of the mass segregation degree when using the MST method.~This is illustrated in Figure~\ref{fig:methods_compare_nbr}, where we see that the mass segregation degree for ordered pairing functions is biased toward very high values for relatively small massive-star samples. These extremely high degrees of mass segregation are not physical, because they are mainly produced by the short MST edges in the binary systems. This is illustrated by the sharp steps in the solid curves, which greatly deviates from the non-mass-segregation distribution equal to unity. Therefore, the gmMST and mMST methods should not be considered as a global compactness measure of massive stars in cases where binary fractions are significant, i.e. $>\!10\%$. Massive stars may have much shorter MST edges than low-mass stars, due to the preferential presence of binaries among massive stars \citep[e.g.][]{san14}.

\subsubsection{Distinction between hard and soft binary stars}
\label{ln:hardsoftbins}

\begin{figure*}
	\centering
	\includegraphics[width=0.8\linewidth,angle=0]{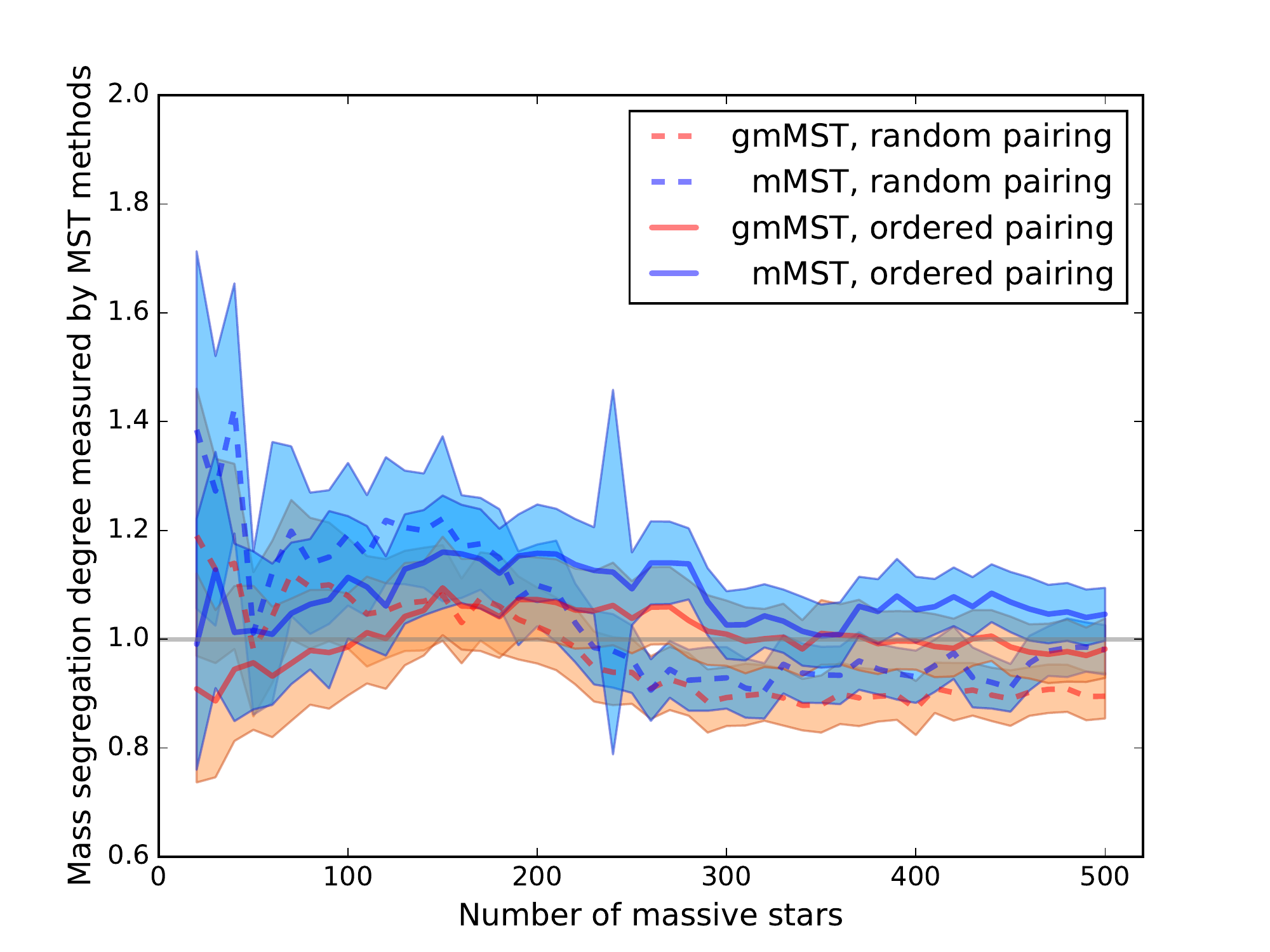}
	\caption{Mass segregation degree based on the geometric-mean MST method (gmMST; red curves) and the median MST method (mMST; blue curves) as a function of the massive-star group size, i.e. the number of most massive stars, for a star cluster with 50\% binaries (for clarity, we do not show star cluster with 10\% binaries). In contrast to the curves in Figure~\ref{fig:methods_compare_nbr}, hard and soft binary stars are treated as single entities or resolved systems, respectively. See Section~\ref{ln:hardsoftbins} for more details. The curves for random binary pairing functions are shown by dashed lines, while ordered pairing functions are given by solid lines. Note that all the curves for random and ordered binary pairing functions are now close to unity (horizontal gray line), as implemented in the model setup. The shaded bands give the corresponding measured 25th to 75th percentiles.}
	\label{fig:methods_compare_wbr}
\end{figure*}

\begin{figure*}
	\centering
	\includegraphics[width=0.75\linewidth,angle=0,trim=0 0 0 10]{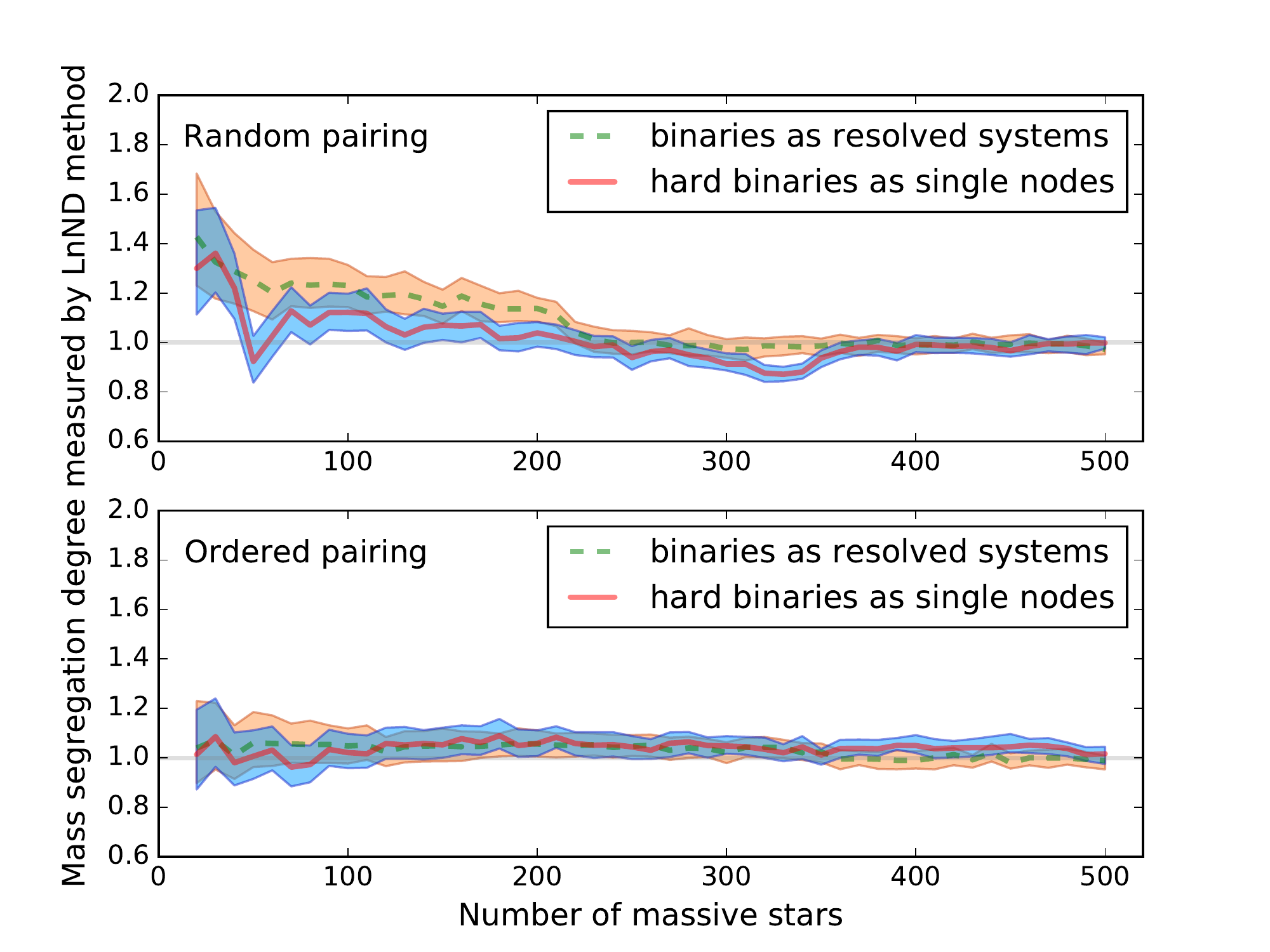}
	\caption{Mass segregation degree based on the LnND method as a function of the massive-star group size, i.e.~the number of most massive stars. The model contains 50\% binaries, while the binaries follow the random pairing function (top panel) and the ordered pairing function (bottom panel). The curves for that hard binary star are treated as single entities and are shown by red solid lines, while all binary stars are treated as resolved systems and are shown by green dashed lines. The shaded bands give the corresponding measured 25th to 75th percentiles. Note that all the curves for random and ordered binary pairing function are close to each other and close to unity (horizontal gray line), as implemented in the model setup.}
	\label{fig:binary_impact_lsd}
\end{figure*}

In contrast, the LnND method is less affected by a high fraction of binaries. The LnND method utilizes the $r_6$ distance, which already reduces the impact of binaries, since the binaries bias mostly the measurement of the nearest distances. The comparison of the impact of binaries on the degree of mass segregation measurements by the LnND method is shown in Figure~\ref{fig:binary_impact_lsd}.

One natural way to avoid this bias is to reduce the weight of the MST edges that connect binary stars. However, instead of introducing a weighting scheme, we treat hard binaries as a single node that is equivalent to other single stars when calculating the degree of mass segregation. According to the Heggie--Hills Law \citep{heg75, hil75}, hard binaries harden and heat a cluster. They hereby behave as a single dynamical unit, in the sense that the very short MST edge within the binary remains in the binary. Therefore, we set a binary system with binding energy greater than the mean kinetic energy of cluster stars to be a ``single star'' instead of two stars, and then we use the MST and LnND methods to calculate the degree of mass segregation. 

We apply this procedure to all binaries, including massive binaries and low-mass binaries, before implementing the gmMST and mMST methods. The new results are shown in Figure~\ref{fig:methods_compare_wbr}, which clearly demonstrates the correct degree of the low-level mass segregation, as implemented in the models. For the subsequent analyses in this paper, we choose the median MST edge (mMST) method as our measurement of compactness, because it is least influenced by extreme values.

\cite{san14} found all massive-star binary systems to be dynamically hard. To test our treatment of binaries in such an extreme case, when there is a large fraction of ``hard'' binaries, we use the same model including 50\% binaries, but now with all binaries manually assigned to be dynamically ``hard.'' For this, we do not use the period distribution as described in Equation~\ref{eqn:bin_period}, but we manually set the period of all the binaries to be $P = 10^{5} P_{\rm min}$, so that the separation of all the binaries is very small.

\begin{figure*}
	\centering
	\includegraphics[width=0.75\linewidth,angle=0,trim=0 0 0 10]{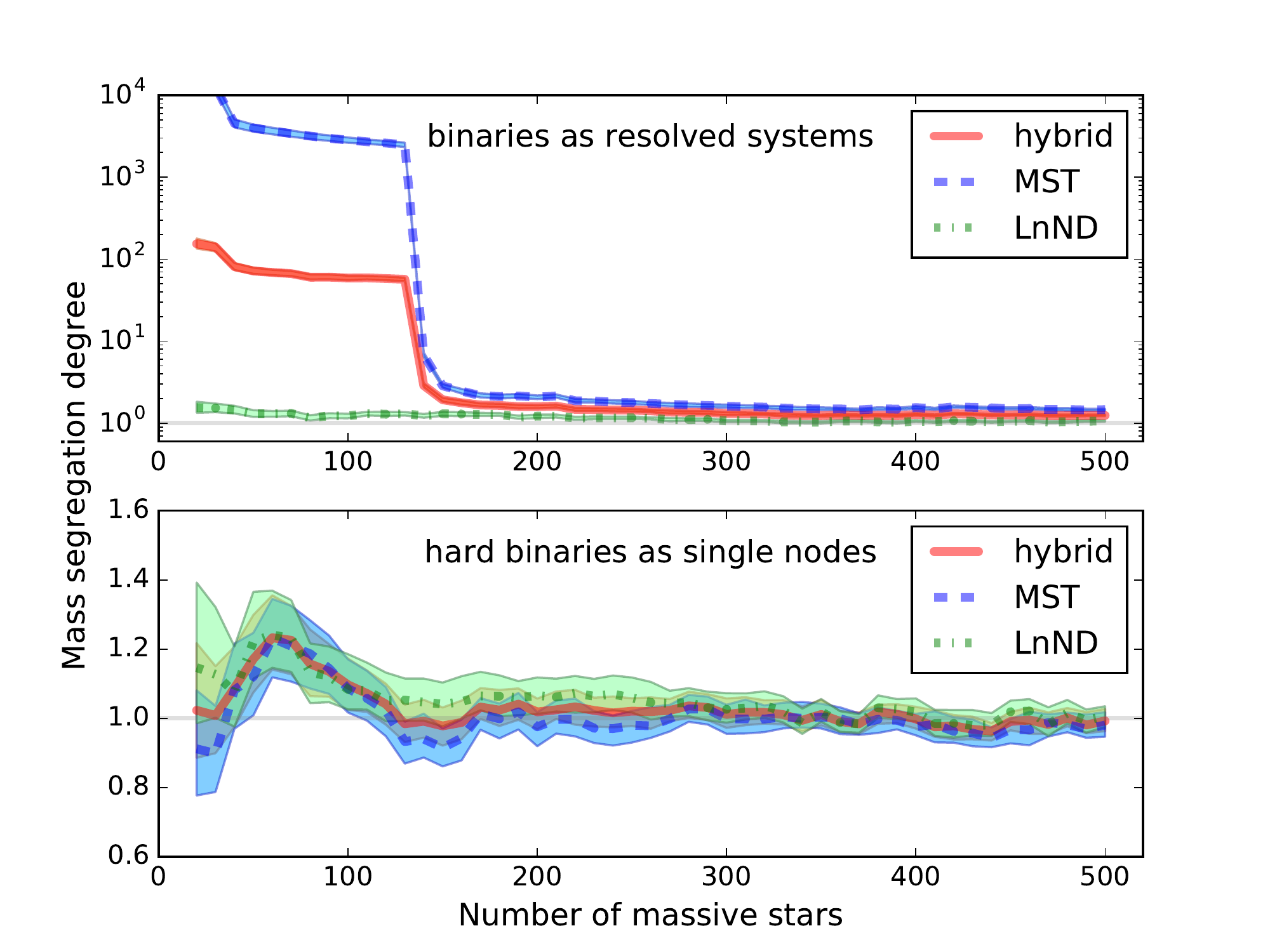}
	\caption{Mass segregation degree based on the hybrid method (hybrid; red solid curves), MST method (MST; blue dashed curves) and LnND method (LnND; green dash-dotted curves), as a function of the massive-star group size, i.e. the number of most massive stars. The shaded bands give the corresponding measured 25th to 75th percentiles. The model contains 50\% binaries, with all the binaries being dynamically hard. The mass segregation measured by different methods when all the binaries are treated as resolved systems is shown in the top panel, while the hard binaries are treated as single nodes and are shown in the bottom panel. When hard binaries are treated as single nodes (bottom panel), all methods give a mass segregation degree close to unity (horizontal gray line), as implemented in the model setup.}
	\label{fig:hard_binary}
\end{figure*}

If the hard binaries are treated as single nodes in our model, their impact on the mass segregation measurements is negligible, as shown in the bottom panel of Figure~\ref{fig:hard_binary}. It is, therefore, essential to identify and exclude the hard binary systems from the MST edge for a reliable measurement of the overall mass segregation in observed star clusters.

\section{Discussion}
\label{sec:discussion}

\subsection{Mass segregation in a simulation of a substructured star cluster}
\label{sec:simulation}

We test these methods with one of our $N$-body simulations. We use a dynamically cold\footnote{This corresponds to the virial ratio of $Q=0.25$, where $Q=T/U$, which is the ratio of kinetic ($T$) to potential energy ($U$).} and fractally distributed star cluster of 100,000 stars as the initial condition, as shown in Figure~\ref{fig:init}. The star cluster is generated to have a fractal dimension of 2.0 \citep[see][for more details]{goo04}. The star cluster is initially highly substructured, and the subclusters are rapidly merging and erase the originally fractal structures throughout its first few megayears of evolution \citep{Fujii2012}. We also include 50\% binaries with the orbit distribution and ordered pairing function, as discussed in Section~\ref{sec:binary_impact}. We adopt the IMF from \cite{kro02} within the stellar mass range $0.08\!-\!100\,M_\odot$ and follow the stellar evolution self-consistently. The masses of the stars are randomly assigned so that the star cluster is set to be non-mass-segregated. The simulation is carried out with {\sc Nbody6} \citep{aar03}.

\begin{figure*}
	\centering
	\includegraphics[width=0.8\linewidth,angle=0,trim=0 0 0 15]{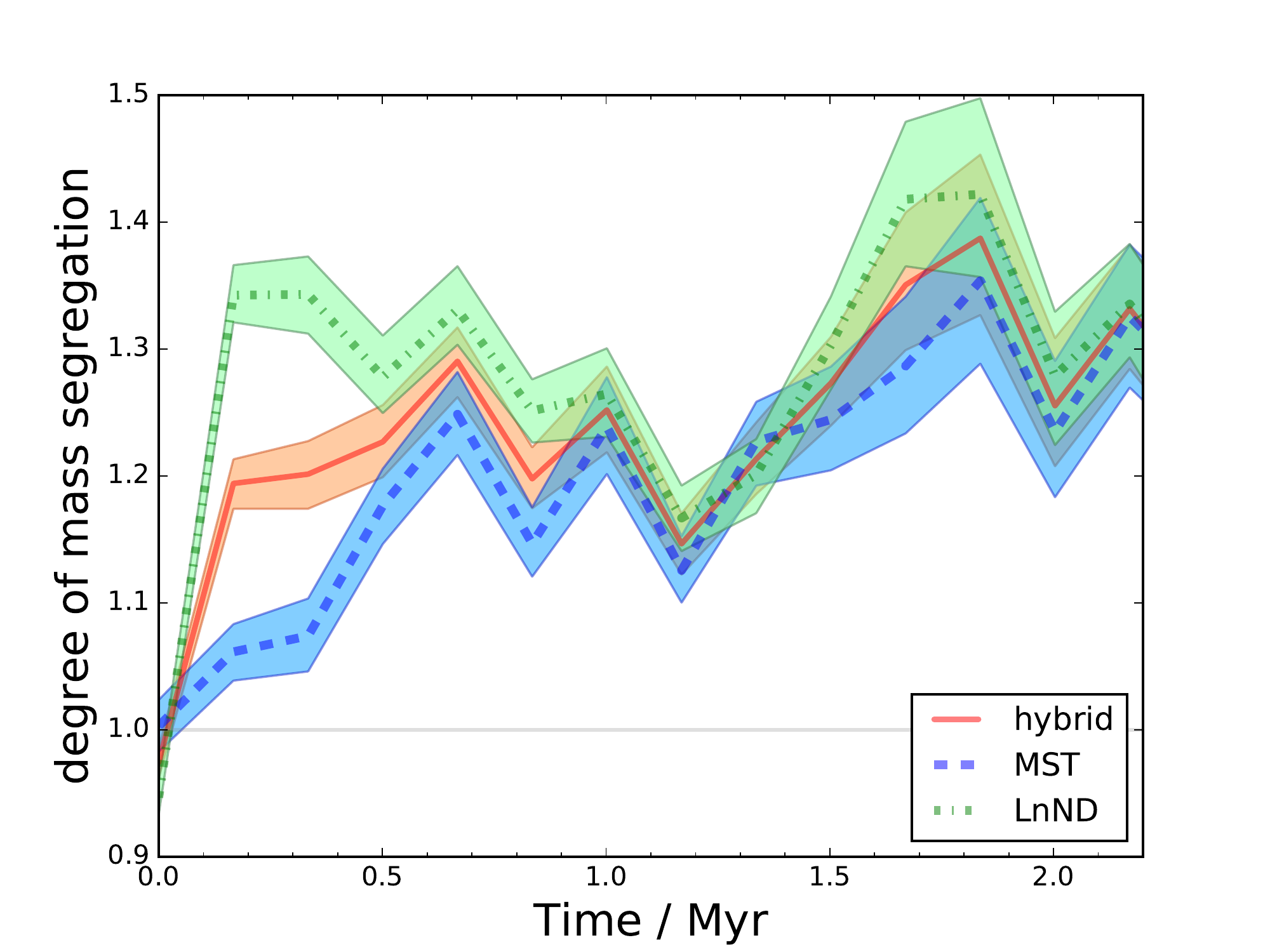}
	\caption{Comparison of different measurements, our hybrid method (hybrid; red solid curve), the MST method (MST; blue dashed curve), and the LnND method (LnND; green dot-dashed curve), using our simulation test. The shaded bands give the 25th to 75th percentiles of the corresponding measurements. Generally, they are compatible with each other once the substructure disappears at approximately $1$\,Myr. Note that the green curve (LnND) is steeper than the blue curve (MST) in the first 0.5 Myr, which indicates that the growth of mass segregation is dominated by the increase of mass segregation inside the subclusters.}
	\label{fig:simulation_test}
\end{figure*}

The comparison of the different segregation measurement methods, which are performed in 3D for the 500 most massive stars, is shown in Figure~\ref{fig:simulation_test}. The star cluster mass-segregates rapidly within $\sim\!2$\,Myr. The various mass segregation measurement methods show some discrepancies during this early evolution, which can be understood from our earlier analysis. The LnND method gives a steeper curve in the first 0.5 Myr during the most vigorous merging period of the subclusters. At this stage, the growth of the mass segregation is dominated by the increase of mass segregation contributed by crowdedness. The massive stars sink into the local overdensities, forming highly mass-segregated subclusters, corresponding to high local mass segregation. On the other hand, the subclusters are relatively far away from each other, corresponding to low mass segregation on a global scale. Considering the mass segregation only locally or globally is not enough to determine the dynamical status of the star cluster during this violent epoch.

Comparing the star cluster at 0.4 and 1.4 Myr, we observe differences in the MST and the LnND value, while the hybrid method gives very similar values. This implies that massive stars are closer to each other, while the subclusters are less dense at 1.4 Myr. However, if we only compare these two states, we are unable to determine which state is more mass segregated, or which is dynamically older. One possibility is that the subclusters of the star cluster become denser and then merge into larger entities. Alternatively, the subclusters may merge into large but loose entities and then become denser. If we treat both scenarios as valid, the hybrid method, which mixes the compactness indicators and the crowdedness indicators, is a reasonable quantity to describe the mass segregation degree, although the discrepancy information is lost. This demonstrates the importance of measuring the mass segregation via both compactness and crowdedness simultaneously.

\subsection{Mass segregation of ONC}
\label{sec:ONC}

\begin{figure*}
	\centering
	\includegraphics[width=0.8\linewidth,angle=0,trim=0 0 0 15]{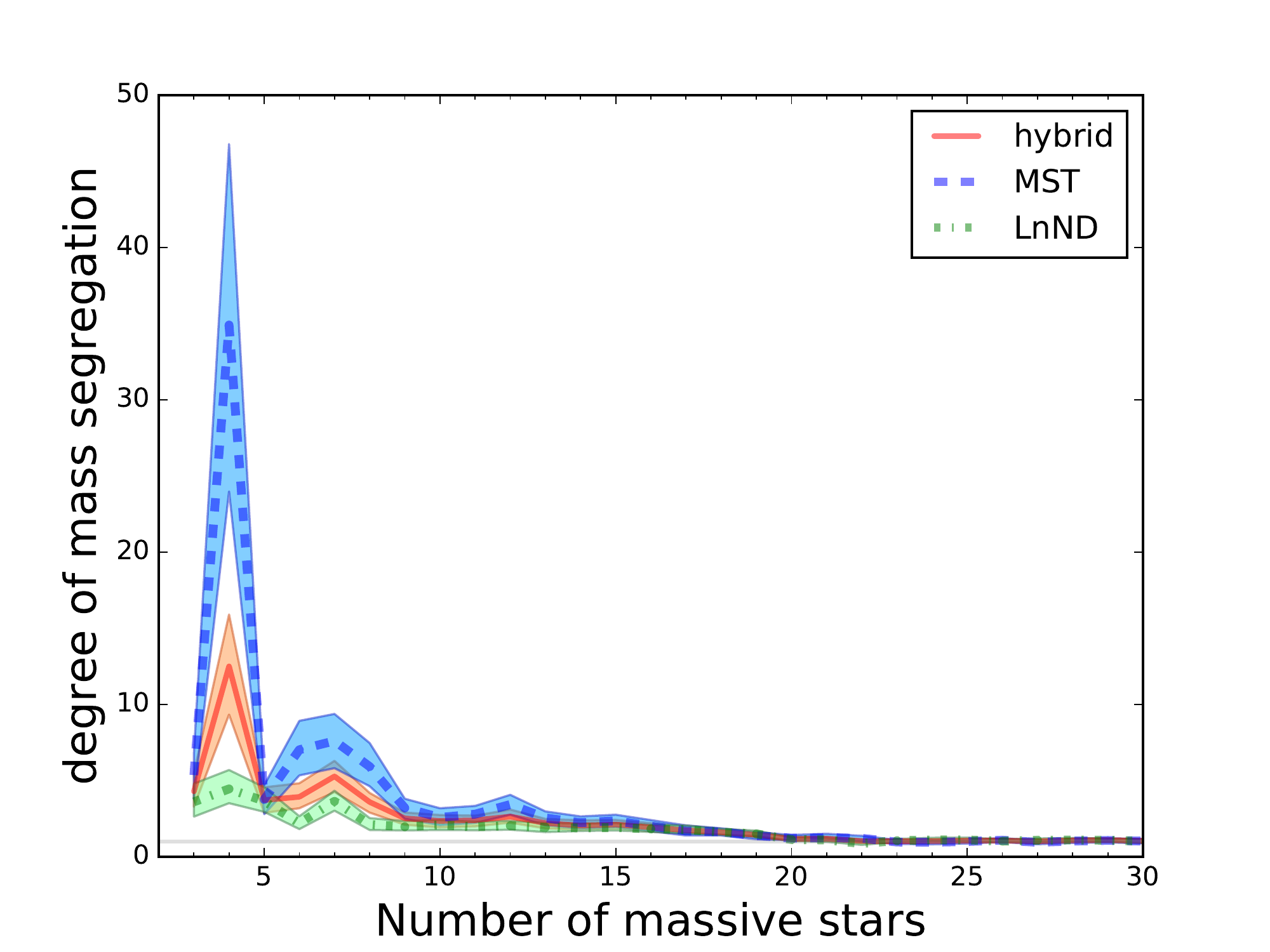}
	\caption{Comparison of different mass segregation measurement methods, including our hybrid method (hybrid; red solid curve), the MST method (MST; blue dashed curve), and the LnND method (LnND; green dot-dashed curve), using ONC data. The shaded bands give the 25th to 75th percentiles for each indicator. The extremely high degree of mass segregation provided by the MST method for the four most massive stars is due to the ONC trapezium.}
	\label{fig:ONC_test}
\end{figure*}

We apply the previous methods to the observational data of the Orion Nebula Cluster (ONC), obtained by \cite{hil97}. We select 929 stars with estimated masses out of a total of 1576 stars. The result is illustrated in Figure~\ref{fig:ONC_test} and clearly shows the mass segregation of the ONC. Given that the ONC is very young, $\sim\!1$\,Myr, only very massive stars (i.e. the 20 most massive stars) are clearly segregated. We also find the degree of mass segregation to be extremely high for the four most massive stars, with MST values reaching as high as $\sim\!35$; these stars are forming the well-known trapezium configuration \citep[e.g.][]{mcc94, hil98}. In contrast, the degree of mass segregation given by the LnND method for these four massive stars is relatively low, but significantly above unity, suggesting that the density around the trapezium is similar to other massive stars in the cluster (e.g.~the fifth or seventh most massive star). 

However, we are unable to determine whether the trapezium-like configuration is accidental or formed in situ by stochastic star formation processes and/or dynamical evolution only by investigating the mass segregation indicators. If the trapezium is purely accidental, the degree of mass segregation should be similar to that of the lower-mass stars, i.e.~the fifth or seventh most massive star. If true, in this case, the extra compactness should not be considered. If, on the other hand, it is formed in situ or through dynamical evolution, the mass segregation can be considered ``real,'' given that the massive stars are closer to each other. If this case is true, the compactness should be fully taken into account in the determination of the mass segregation estimate. Unfortunately, this knowledge is not available to the observer a priori. Therefore, we consider both assumptions as potentially valid by giving balanced information on the degree of mass segregation using the hybrid method. This is our primary goal in advocating the usage of the hybrid method.

We conclude that the ONC is significantly mass segregated for the 20 most massive stars. In Figure~\ref{fig:ONC_test}, this corresponds to the $m$ number of stars for which all mass segregation indicators reach the unity level. The extreme value down to the fourth most massive star is caused by the trapezium-like configuration.

\subsection{Mass segregation of Taurus}
\label{sec:Taurus}

\begin{figure*}
	\centering
	\includegraphics[width=0.8\linewidth,angle=0,trim=0 0 0 20]{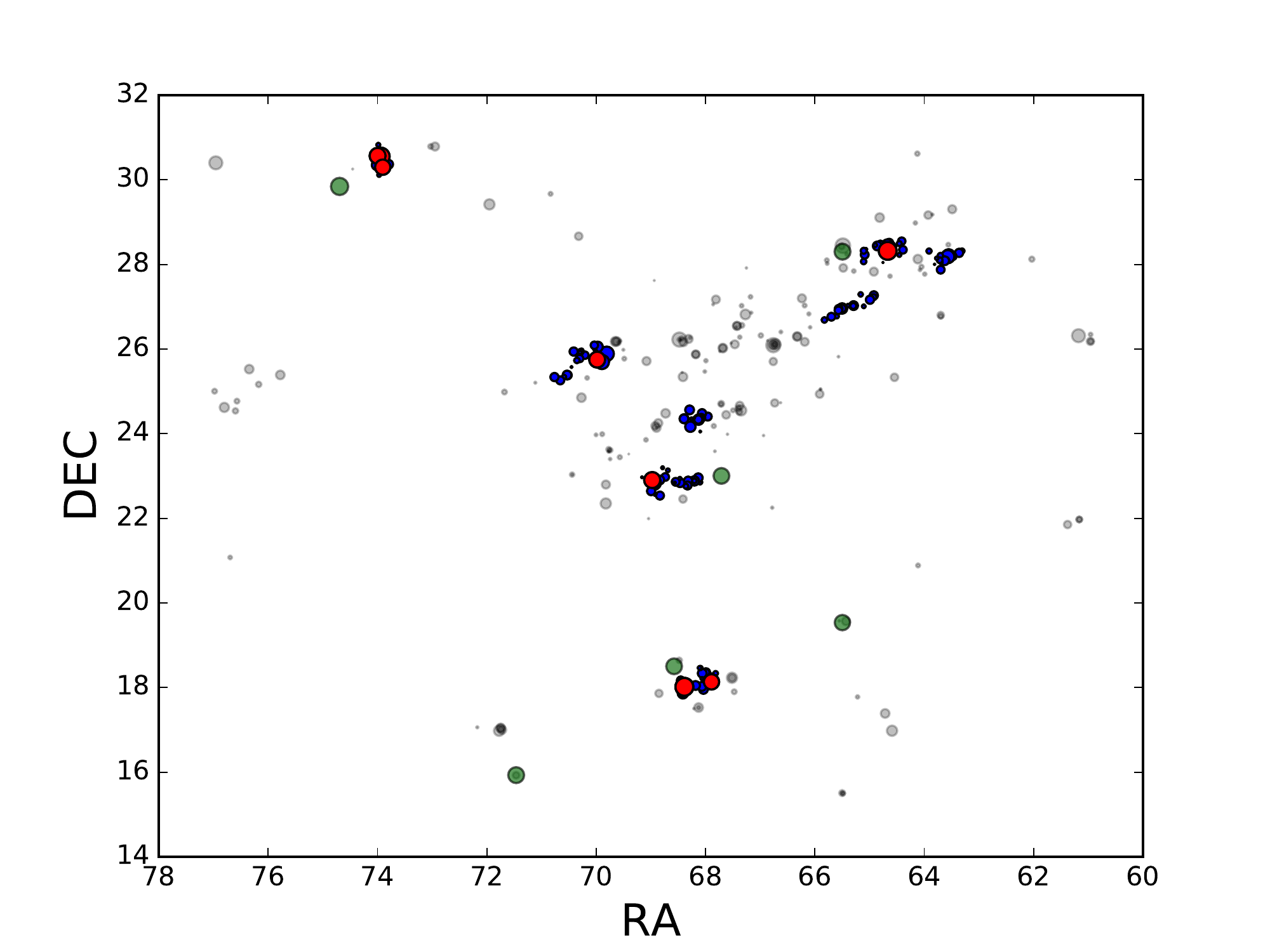}
	\caption{Illustration of the stars of Taurus. The size of the circle represents the mass of the stars. Eight groups are identified following the method of \cite{kir11}. The member stars belonging to these eight groups are shown in blue, while the nonmember stars are shown in black. The fifteen most massive stars are overplotted in red (group members) and green (nongroup members).}
	\label{fig:Taurus}
\end{figure*}

Taurus is a young cluster ($\sim\!1$\,Myr; \cite{ken94}) with multiple clumps. \cite{gom93} identified six groups using the simple grid technique and the kernel method. \cite{kir11} identified eight groups by pruning the long MST edges. We adopt the data of \cite{kir11}, which contain 352 stars with estimated masses in total. The whole cluster is identified as eight sparse groups containing 178 member stars. As shown in Figure~\ref{fig:Taurus}, the group members are shown in blue, while the nongroup members are shown in black. The fifteen most massive stars are overplotted in red (group members) and green (nongroup members).

\begin{figure*}
	\centering
	\includegraphics[width=1.1\linewidth,angle=0,trim=50 0 0 0]{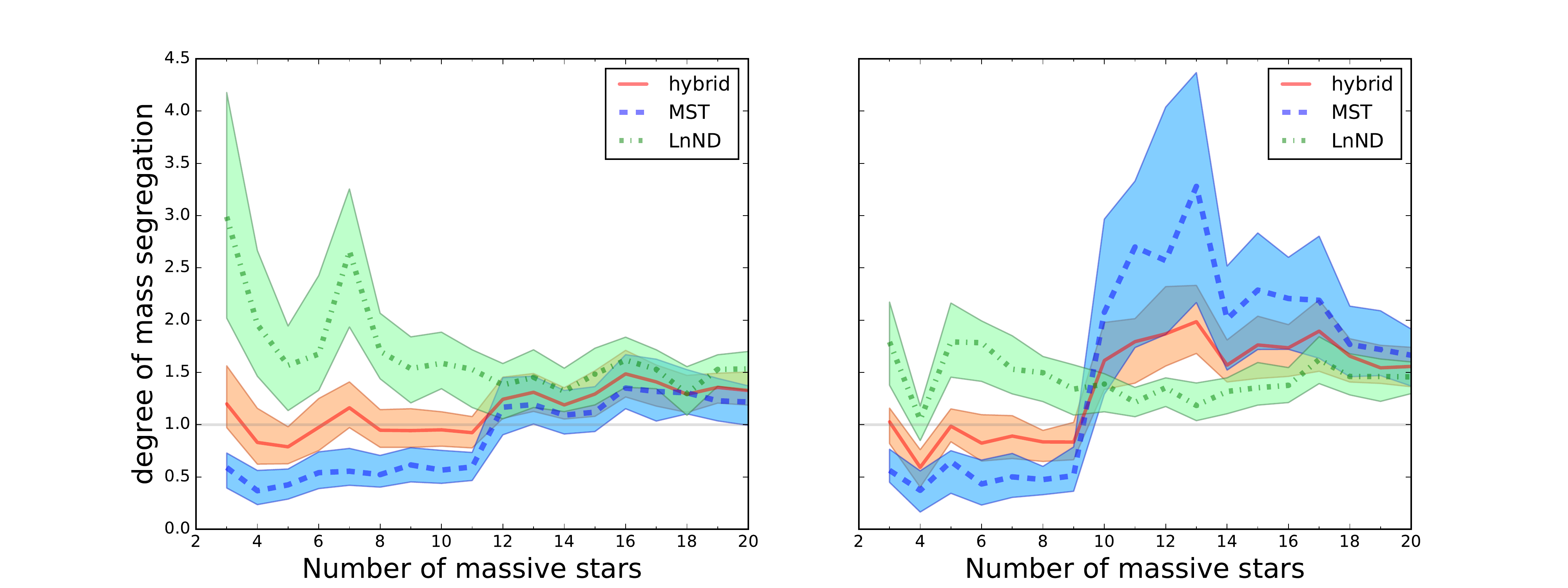}
	\caption{Comparison of different mass segregation measurement methods, including our hybrid method (hybrid; red solid curve), the MST method (MST; blue dashed curve), and the LnND method (LnND; green dash dot curve), using Taurus data. The shaded bands give the 25th to 75th percentiles for each indicator. Left panel: mass segregation of all 352 stars in Taurus. Right panel: mass segregation of 178 stars that are identified as members of eight groups in Taurus \citep{kir11}.}
	\label{fig:Taurus_test}
\end{figure*}

The result of mass segregation using our approaches is illustrated in Figure~\ref{fig:Taurus_test}. The left panel shows the mass segregation of all 352 stars of Taurus, while the right panel shows the mass segregation of 178 stars that are identified as group members. When the number of massive stars is small, the LnND method and the MST method give opposite results of the mass segregation of the cluster, while the hybrid method gives a close-to-unity value of mass segregation degree. The green dot-dashed curve given by the LnND method indicates that the star cluster is significantly mass segregated down to the 20th most massive stars. The blue dashed curve representing the MST method shows a turning point. When the number of massive stars is less than 11 (left panel) or nine (right panel), the star cluster shows ``inverse'' mass segregation using the MST method.

\begin{figure*}
	\centering
	\includegraphics[width=0.8\linewidth,angle=0]{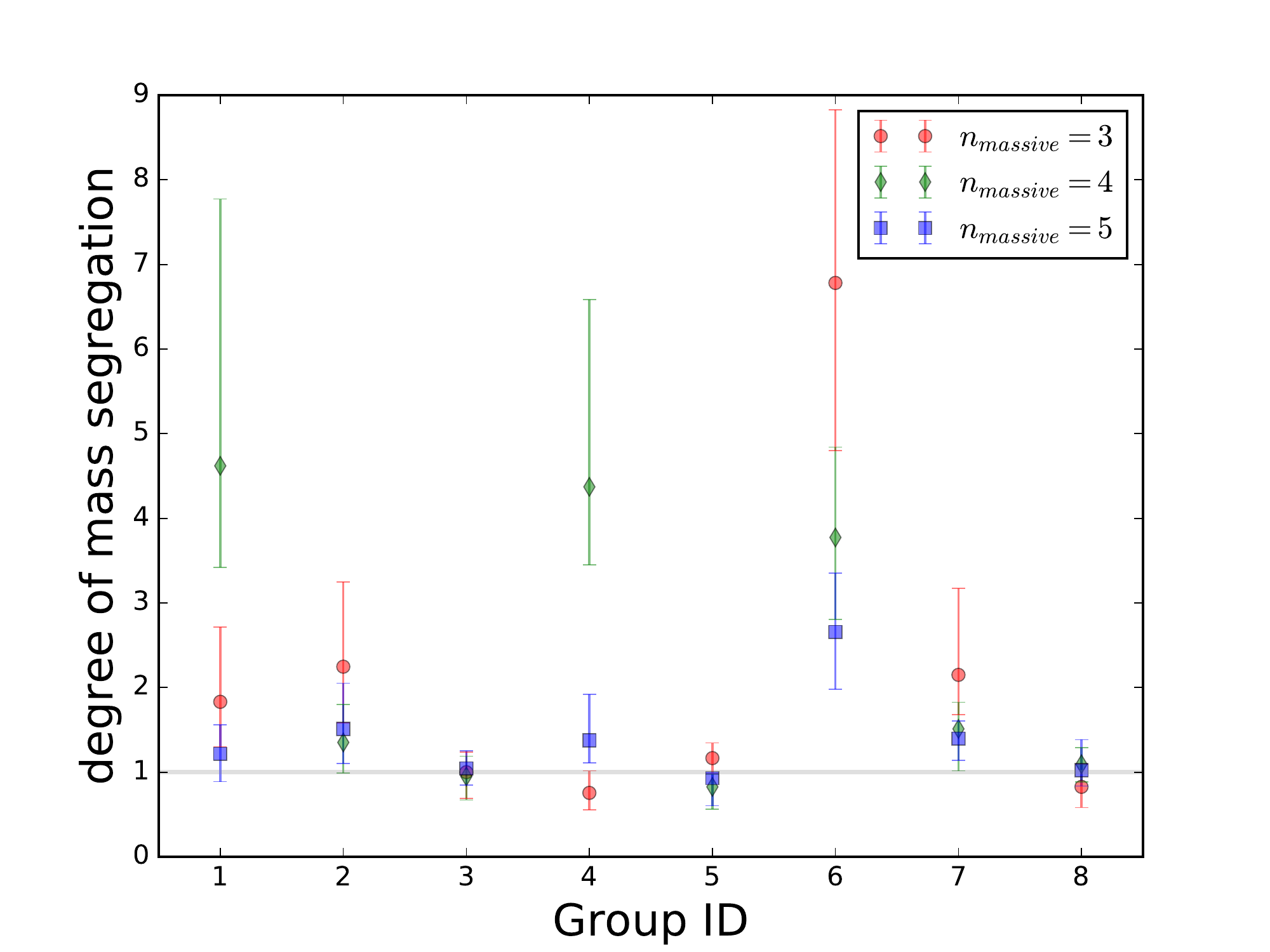}
	\caption{Mass segregation degree based on the MST method for the three most massive stars (red circles), four most massive stars (green diamonds), and five most massive stars (blue squares) of each individual group of Taurus. The error bars give the 25th to 75th percentiles for each measurement.}
	\label{fig:Taurus_sub_test}
\end{figure*}

We notice that from the whole star cluster scale, the most massive stars distributed sparsely. If no dynamical evolution is taken into account, the massive stars do not form as a group, but form independently of the molecular clouds where they are located. On the other hand, from the molecular cloud scale, or from the scale of the group, massive stars tend to be distributed in the center of their groups \citep{kir11} and have high crowdedness. Furthermore, for most of the groups, their massive members are tend to be compact, as shown in Figure~\ref{fig:Taurus_sub_test}. Therefore, the Taurus is not as highly mass segregated at the whole cluster scale as it is locally. Neither is it accurate to describe the cluster as ``inversely'' mass segregated even at the whole cluster scale; otherwise, the massive stars should only be found in the outer groups. This is why we give a close-to-unity value with the hybrid method by averaging both the compactness indicator and the crowdedness indicator.

The MST method compares the compactness of massive stars and that of the same number of randomly chosen star sets. When the number of massive stars is small, the compactness of the massive stars is dominated by the large separations between the groups since they distributed sparsely in many different groups. As a comparison, a randomly chosen star set has a larger chance of being distributed in fewer groups; hence, the corresponding compactness is of the scale of the group size. Therefore, given that the massive stars do not show any preference of being located in the inner groups or the outer groups, the reason that Taurus shows ``inverse'' mass segregation indicated by the MST method is that the number of massive stars is not large enough compared to the number of subclusters, or the number of groups. This is also the reason that the turning point in the left panel is larger than that in the right panel, since there are more ``clumps'' when considering all 352 stars. If the number of massive stars increases, the compactness gets higher. As a consequence, Taurus shows significant mass segregation for the 15 most massive stars by both the LnND method and the MST method.

The MST method is less efficient in discussing the local mass segregation when the number of massive stars is small and when the star cluster is highly substructured like Taurus. However, it still provides the information that massive stars are not formed in one clump but can be formed in any sparsely distributed molecular clouds. If the local mass segregation, or the mass segregation at subcluster scale, is of interest, the LnND method better describes the status of the star cluster. Otherwise, the measurements should be applied for each individual clump after identifying the clumps of the whole star cluster. If, on the other hand, the global mass segregation, or the mass segregation at the whole star cluster scale, is of interest, the compactness measured by the MST method should be taken into account. Therefore, the crowdedness and the compactness should be measured simultaneously in order to better describe the mass segregation of the whole cluster.

We conclude that Taurus is significantly mass segregated for the 20 most massive stars locally, or at subcluster scale, while at the whole cluster scale, it is not as highly mass segregated as it is locally. We also find that the massive stars tend to form in the overdensity regions of many different clumps rather than in one clump.

\section{Conclusions}
\label{sec:conclusions}

We have developed a new method of measuring the degree of mass segregation by combining the MST approach with the LnND technique. Our method inherits all the advantages provided by both MST- and LnND-based methods: (1) a quantitative value so that different star clusters are comparable, (2) being independent of star cluster geometry, and (3) imposing no requirement of defining the star cluster center.

We also distinguish compactness and crowdedness of massive stars. The compactness of a group of stars measured by the MST method focuses more on the global mass segregation, while the crowdedness measured by the LnND method focuses more on the local mass segregation. We show that providing only the compactness or the crowdedness is not enough in investigating the mass segregation of star clusters. From the star formation point of view, if massive stars are formed in one group, then they have high compactness with relatively low crowdedness. If, on the other hand, massive stars are formed in the overdensity regions of many different subclusters, then they have high crowdedness with relatively low compactness (see Section~\ref{sec:Taurus}). Measuring both compactness and crowdedness enables us to investigate the possibilities of the formation of massive stars. From the dynamical evolution point of view, if the subclusters become denser before violent merging, then massive stars sink into the center of the subclusters; hence, high crowdedness but relatively low compactness can be detected (see Section~\ref{sec:simulation}). If, on the other hand, the merging of subclusters is predominant before their collapse, then massive stars are gathered together at the whole star cluster scale without changing much of their local density; hence, high compactness but relatively low crowdedness can be detected.

We therefore suggest to measure the local and global mass segregation simultaneously. In addition, if the unmatched local and global mass segregation is found and can be explained by two possible assumptions, e.g.,~if the high compactness is purely accidental or caused by dynamical evolution, the hybrid method that combines these two indicators should be considered when both assumptions are treated as valid. This is why we propose our hybrid method, which enables us to track the aspects of mass segregation from the perspective of both the global and the local configuration (see Section~\ref{sec:ONC}).

Moreover, we discuss the impact of binaries in measuring the mass segregation. The mass segregation measured by the MST method is greatly biased when the binary fraction, especially the close binary fraction, is high in a star cluster, while the LnND method is much less affected. We treat the hard binaries as single nodes in the sense that the short MST edge within the binary remains in the binary. By doing so, we successfully eliminate the impact of binaries on the mass segregation parameter. This, however, implies that an accurate knowledge of the fraction of hard binaries is required for robustly measuring the mass segregation in observed star clusters.

\section*{Acknowledgments}
J.Y. gratefully acknowledges support in the form of a FONDECYT Postdoctoral Fellowship (no.~3130646). T.H.P. acknowledges support through FONDECYT Regular Grants (no.~1121005 and no.~1161817) and the BASAL Center for Astrophysics and Associated Technologies (PFB-06). We thank Michael Fellhauer, Richad de Grijs, and the members of the Complex Stellar Systems team at IA-PUC for stimulating discussions.

\begin{appendix}
\section{Calculation of the MST}

\subsection{Acceleration via Delaunay triangulation and $k$-d tree}
Computing the MST value directly is somewhat computationally expensive on generic desktop computers. However, as suggested by \cite{olc11}, a Delaunay triangulation can be used to accelerate constructing the MST for a projected star cluster, i.e.~in two dimensions. In Euclidean space, the MST is a subgraph of the Delaunay triangulation \citep{pre85}. We use the {\sc Triangle} software package\footnote{Code is available from: \urlstyle{rm}\url{https://www.cs.cmu.edu/~quake/triangle.html}} \citep{she96} to construct the Delaunay triangles and then use the {\sc Kruskal} algorithm to compute the corresponding MST. The Delaunay triangulation can be constructed in a time that scales as $\mathcal{O}\bigl(N \log N \bigr)$, while the {\sc Kruskal} algorithm also costs $\mathcal{O}\bigl(N \log N \bigr)$ of computational time, where $N$ is the number of points, i.e.~stars in the sample. The median edge of the MST can be selected in $\mathcal{O}\bigl(\log N \bigr)$ time with a generic quick-select routine. Therefore, the total computational timescales are $\mathcal{O}\bigl(N \log N \bigr)$. In addition, we set the number of random sets to be $k$, so that the fraction of stars that are not chosen from the entire star cluster of size $N$ is
\begin{equation}
	p = \left( \frac{ N - m } { N } \right)^k \quad \Rightarrow \quad k = \lceil{ \frac{ \ln( p ) }{ \ln( 1 - m/N ) } }\rceil \,,
\end{equation}
where $m$ is the random sample size and $\lceil{x}\rceil$ is the ceiling function of $x$. We set $p = 0.01$ following \cite{olc11}.

In the 3D case, there is no algorithm that can construct the MST in $\mathcal{O}\bigl(N \log N \bigr)$ time. However, a nearest-neighbor-based {\sc Prim} scheme \citep{ben78} can be implemented in $\mathcal{O}\bigl(N \log N \bigr)$ expected time, which can be easily extended to higher dimensions. The {\sc Prim} algorithm can be described as follows: first choose one arbitrary vertex as the starting tree with only one point, and then grow the tree by finding a vertex from the rest of the vertices that has the shortest distance to the tree; repeat this process until all vertices are in the tree. This algorithm takes $\mathcal{O}\bigl(N^2 \bigr)$ time if brute-force searching is used to find the shortest distance point. Instead of traversing all vertices, the nearest-neighbor-based scheme uses a $k$-nearest neighbor (KNN) scheme (here $k\!=\!1$), which can find the nearest vertex in $\mathcal{O}\bigl(N \log N \bigr)$ expected time. Therefore, the total computational time is expected to be $\mathcal{O}\bigl(N \log N \bigr)$, and we can use Delaunay triangulation and KNN to measure the compactness of a group of massive stars in a very efficient way.

\end{appendix}

\end{document}